# New Property Averaging Scheme for Volume of Fluid Method for Two-Phase Flows with Large Viscosity Ratios

**Sucharitha Rajendran** (ASME Member)

Thermal-Fluids and Thermal Processing Laboratory,

Department of Mechanical and Materials Engineering,

University of Cincinnati, Cincinnati, OH 45221

**Raj M. Manglik** (ASME Fellow)

Thermal-Fluids and Thermal Processing Laboratory,

Department of Mechanical and Materials Engineering,

University of Cincinnati, Cincinnati, OH 45221

**Milind A. Jog**[1] (ASME Fellow)

Thermal-Fluids and Thermal Processing Laboratory,

Department of Mechanical and Materials Engineering,

University of Cincinnati, Cincinnati, OH 45221

---

[1] Corresponding author. Email: Milind.Jog@uc.edu

**Abstract**

To predict liquid-gas two-phase flow phenomena, accurate tracking and prediction of the evolving liquid-gas interface is required. Volume-of-Fluid or VoF method has been used in the literature for computationally modeling of such flows. In the VoF method, a single set of governing equations are solved for both phases along with an advection equation for the volume fraction. The properties in each computational cell are determined by a linear weighted average of the properties of the two fluids based on the phase fraction. While the method predicts water-air flows well, the predictions tend to deviate significantly from experimental data for liquids with high viscosity. A new property averaging technique is proposed in this paper, which is shown to provide accurate results for high viscosity liquids. Computational predictions using the open source VoF solver *interFoam* (available as a part of the OpenFOAM computational tool), and those obtained using the proposed method are compared with experimental data for multiple two-phase applications. Four different problems, viz., suspended droplet in air, jet breakup, drop impact on thin films, and air entrapment during drop interaction with liquid pool, are considered to extensively validate the new method. Experimental data for water and aqueous solutions of propylene and ethylene glycol are used to cover a range of surface tension (72 – 36 mN/m) and viscosities (1 – 40 mPa.s). For all cases, the modified VoF solver is observed to perform significantly better than original VoF method. It reduces any spurious currents in simulations of drop suspended in air. For the cases of drop impacting on a pool and during drop generation from liquid jets, the time progression of the surface tension governed dynamics is improved from the slower estimate of *interFoam* solver. In the case of drop impacting on a thin liquid film, where it is critical to capture the intricate interplay between the surface tension and viscous force, the former effect of surface tension is exaggerated in the original method but correctly captured with the proposed new method.

**Introduction**

Our colleague of many years, Professor Kirti "Karman" Ghia, was devoted to the advancement of fundamentals and applications of computational fluid dynamics. An example is his early work on the lid driven cavity flow [1] that made a seminal contribution to multigrid methods and became a benchmark for work that followed (over 5000 citations to date!). We enjoyed many technical discussions with him. This work is a tribute to Prof. Ghia, and it addresses a limitation of the Volume-of-Fluid (VoF) method in accurately predicting two-phase flows when the viscosity ratio of the two fluids is large.

Two-phase processes with jets, droplets, and bubbles such as droplet impact on dry surfaces, or on thin films, or on liquid pools, jet breakup and drop formation, and bubble growth from an orifice submerged in liquid, are encountered in a variety of industrial applications [2]. Numerical modeling of such phenomena has steadily advanced, and along with experiments, it has provided insights into the underlying physics of the processes. For example, Berberovic et al. [3] used numerical techniques along with experimental observations to understand the dynamics of drop impact on liquid surfaces and have characterized formation and propagation of the surface wave during impact. Extensive investigation of dynamics of confined bubbles in laminar micro channel flow by Khodaparast et al. [4] relies on experimental and numerical analysis of two-phase flow. Similarly, research in jets and atomization characterization [5–7] has advanced with the help of numerical techniques. In addition to flow characterization, coupled heat transfer solution has been implemented by Trujillo et al. [8] to study drop impact heat transfer. Sanjivan et al. [9] have incorporated interfacial surfactant adsorption/desorption kinetics in their computational simulations of bubble growth in aqueous surfactant solutions. These and many such other studies have used numerical methods for predicting multiphase (or two-phase) flow behavior.

For accurate predictions of the two-phase phenomena encountered in applications associated with liquid breakup, atomization, liquid collisions and entrainment, precise capture of the liquid and gas interface is essential. Interfacial tracking methods in incompressible fluids include front tracking [10–12] , boundary integral [13], volume of fluid (VoF) [14] and level set (LS) [15,16] methods. In the past decade, the techniques that have received the most attention are the last two – VoF and LS – due to their robustness and ease of implementation. As the interface evolves in a fixed computational mesh, both these methods use a separate parameter to identify the different phases in each cell in the domain. The primary difference in these methods is their treatment of this phase parameter. In the VoF method, an advection equation governs the fraction of volume of the primary phase as the interface moves in time and space. This makes the interface discontinuous and therefore, reconstruction of the calculated interface is required to obtain a smooth curve for curvature and normal estimations at each interface location. The interface is typically smeared over a few cells owing to the change in volume fraction as it goes from one phase to the other. In contrast, the LS method defines the interface by a smooth continuous function going from positive in one fluid to negative in the other and having a value of zero at the interface. This eliminates the requirement for reconstruction and creates an interface solution that is sharp unlike in the VoF method. However, though accurate estimations of the interface topology are achievable, the LS method does not inherently conserve mass by the advection step unlike the VoF method. The VoF method has been implemented in many commercial solvers as it inherently conserves mass. This is the method under consideration for the current study. As interface tracking is very sensitive to the reconstruction, VoF technique was seen to yield acceptable results only for certain applications and working liquids and some efforts have been taken to improve the interface tracking for this method. This is discussed later in the next section.

Prior numerical studies of jets and droplets that use VoF techniques have primarily looked at fluids with viscosities similar to water. They have provided useful results with water, hydrocarbon fuels, and refrigerants – liquids with viscosities close to or less than that of water. However, liquids used in chemical, pharmaceutical and process industries can have significantly larger viscosities. In such cases, with large differences in the viscosities of the two fluids, VoF method was found to produce errors in numerical solution. The error in numerical prediction also increases with higher viscosity working fluid. A few studies have coupled the LS and VoF techniques to improve interface tracking while ensuring mass conservation in the domain [17,18]. The improved accuracy of combined LS and VoF (CLSVOF) method can be observed in the work of Ray et al. (2015) [ ] where the problem of liquid drop impact has been modeled. Some other studies have focused on improving dynamic meshing methods along with the VoF implementation to refine the numerical calculations at the interface [7,19]. The current study provides a modification to the VoF method so that accurate predictions can be made for highly viscous liquids. Extensive validation of the suggested modification to the VoF method is conducted with available experimental data for four different physical processes. The VoF method used for comparison in this study is based on the open source two-phase solver present in OpenFOAM®. This package is well suited to handle complex geometries and is easily parallelizable without the limits of licensing. It is based on C++, where equations can be in a form that has a close resemblance to its mathematical equivalent. This VoF interface tracking solver was first implemented in 1999 by Ubbink & Issa [20]. A detailed summary of the original solver has been presented by Deshpande et al. [21]. Modification to the interface reconstruction is implemented to ensure better predictions for viscous working liquids.

**Numerical Procedure**

The numerical procedure to solve the two-phase incompressible flow problems in this study, and as enabled in the open-source code OpenFOAM 2.2.1 is discussed here. The implemented solver available for this is *interFoam*, which, as noted by Deshpande et al. [21], has been reliably employed to solve a number of problems. However, the accuracy of this original solver was noticed to drop when the viscosity of the working fluid is high. This VoF solver uses finite volume discretization of the governing equations on a fixed grid in the domain. All the flow variables are stored in the cell center in the associated finite volume technique. A modified method is proposed herein and its efficacy and relatively more precise validity are demonstrated and highlighted.

**Governing Equations**

The numerical technique adopted in this study considers two immiscible fluids (a gas and a liquid) that are separated by a sharp interface. However, the two fluids are not solved separately with their different properties. Instead, a single set of equations are solved for the entire fluid region with properties differing continuously from one fluid to the other. The two fluid media are distinguished by a phase fraction property that has values between 0 and 1, where 1 is identified as liquid and 0 as gas. The interface is defined by the region where the value of this phase fraction transitions from 0 to 1. Thus, the interface is smeared in such an evaluation and is not sharp. The definition for this phase fraction ($\alpha$) is as follows:

$$\alpha = \frac{\text{Volume of primary phase}}{\text{Total control volume}} \tag{1}$$

$$\text{where, } \alpha = \begin{cases} 1 & \text{in primary fluid (liquid)} \\ 0 < \alpha < 1 & \text{in transition region (interface)} \\ 0 & \text{in secondary fluid (air)} \end{cases}$$

Using this phase fraction identifier for the fluids the respective thermo-physical properties are determined in each computation cell as follows:

$$\varphi = \alpha\, \varphi_1 + (1 - \alpha)\varphi_2 \tag{2}$$

where, $\varphi$ is the property of the fluid and $\alpha$ is the phase fraction in the cell and subscripts 1 and 2 represent liquid and air, respectively. Equation (2) therefore allows the properties to be calculated in each cell at each time instant in the computations of the evolving interface. The transport of this newly defined fluid bulk property (phase fraction, $\alpha$) is further defined by an advection equation,

$$\frac{\partial \alpha}{\partial t} + \nabla \cdot (\bar{v}\alpha) = 0 \tag{3}$$

Additionally, the finite volume solver uses the continuity and momentum equations to computationally obtain the flow field. In this study, an incompressible flow solution is of interest and hence the energy equation is not added.

**Continuity:**

$$\nabla \cdot (\bar{v}) = 0 \tag{4}$$

**Momentum Conservation:**

$$\frac{\partial(\rho\bar{v})}{\partial t} + \nabla \cdot (\rho\bar{v}\,\bar{v}) = -\nabla p + \nabla \cdot \bar{T} + f_s + \rho\bar{g} \tag{5}$$

where $\bar{T}$ is the viscous stress tensor and $f_s$ is the volumetric force due to surface tension. The properties such as density and viscosity in these two equations are evaluated based on the phase fraction ($\alpha$).

Equations (3) – (4) are solved numerically by adopting a control volume discretization. In this process, the surface tension force term in the momentum equation calls for special consideration as it is not inherently a volumetric term like the other components in the equation. One way to resolve this would be to use this force as a boundary condition on the free surface and the surface pressure is obtained by linear interpolation between the surface pressure required and the fluid pressure inside the interface. This requires multiple iterations to obtain surface pressures within some tolerance interval with respect to that at previous time step. After the correction of the surface pressure, the momentum and continuity equations are re-solved before advancing to the next time step. This was the original VoF method suggested by Hirt and Nichols in 1981 [14]. In addition to higher computational time due to the iterative surface pressure correction, this technique required approximate prior knowledge of the interface shape that was to be obtained from upstream and downstream cells.

Brackbill et al. [22] in 1992, proposed a means to resolve the problems associated with this surface treatment by converting the surface tension forces to an equivalent volume force that can be added to the Navier-stokes equation as an additional body force term. Their method called continuum surface force (CSF) calculates this equivalent volumetric surface tension force as:

$$f_s = \sigma \kappa \, \mathbf{n} \, \delta_s \tag{6}$$

where, σ is the surface tension coefficient, κ is the interface curvature, **n** is the surface normal and $\delta_s$ is a Dirac delta function that assists with concentrating this calculation on the interface. In this model, the interface curvature κ is determined as a function of local gradients of the surface normal, **n** which in turn is a function of the phase fraction ($\alpha$) given by

$$\kappa = \nabla \cdot \hat{\mathbf{n}} \text{ and } \mathbf{n} = \nabla \cdot \alpha \tag{7}$$

Tang and Wrobel [23] have shown how this surface tension model can be written in terms of pressure drop across the interface and expressed in terms of a volume force in the momentum equation. They normalize the interface curvature by using the volume averaged density ($\rho$) to obtain the following for the surface tension force:

$$f_s = \sigma \kappa \frac{\rho}{0.5(\rho_1 + \rho_2)} \nabla \cdot \alpha \tag{8}$$

It must be noted that the numerical solution would depend on the estimation of the interface normal, $\hat{\mathbf{n}}$, which in turn, is a function of the phase fraction. Thus, correct estimation of these factors determines the accuracy and performance of the numerical solution.

**Modifications to the VoF Solver**

One significant cause for incorrect estimation of interface curvature and normal stems from deviations in estimations of fluid properties across the smeared interface. Inaccurate interface calculations, in turn, would result in unreliable surface tension force estimation and thus cause failure in predicting correct interface pressure gradients. To resolve this difficulty in interface tracking, some past studies have either looked at correcting the Courant number estimation, such as the case in Beerners et al.[24], or have proposed alternative methods to estimate interface curvature and therefore surface tension force [25], while still others have looked at better meshing techniques to resolve the computational domain in a dynamic manner [26]. Other notable improvements to accurately predict surface tension-dominated flows are provided by Puckett et al. [ ], Popinet & Zelenski [ ], and Gerlach et al. [ ]. In our preliminary analysis of the complication in correct interface estimation, it was found that the relative viscosity of the two phases in question was an underlying source of inaccuracy. Thus, an attempt to correct this is made by proposing a modification to the interface property averaging technique.

Multiple experimental measurements have been used to test a generic exponential property averaging equation that is defined as:

$$\varphi = \varphi_2 + (\varphi_1 - \varphi_2)\alpha^C \quad (9)$$

The specific experimental case employed for this analysis is the dynamic development of the crown on impact of a liquid drop on a thin film of the same liquid [33]. This case was computationally modeled as well, and numerical solutions duplicating the experimental results obtained with several different working liquid properties (Table 1) are used to identify the correct expression for the constant **C** in the exponential property averaging of Eq. (9). The results obtained are then correlated with their relative viscosity ratios and the consequent results are graphed in Figure 1. Thus, a property estimation equation that depends on the viscosity of the two working fluids (1 and 2) is determined and is given by

$$\varphi = \varphi_2 + (\varphi_1 - \varphi_2)\alpha^{1.5(v_1/v_2)+0.75} \quad (10)$$

where, $v$ is the kinematic viscosity of the fluid in consideration.

*Table 1: Properties of the Newtonian liquids used to determine the exponent C in Eqn.(9)*

| Liquids | Density (kg/m$^3$) | Viscosity (Pa.s) | Surface Tension (N/m) |
|---|---|---|---|
| Water | 998.0 | 0.0010 | 0.0728 |
| 25% by volume of Propylene Glycol (25% PG/ 75%Water) | 1007.5 | 0.00255 | 0.0541 |
| 50% by volume of Propylene Glycol (50% PG/ 50% Water) | 1017.0 | 0.0050 | 0.0452 |
| 75% by volume of Propylene Glycol (75% PG/ 25% Water) | 1026.5 | 0.0120 | 0.0411 |
| Ethylene Glycol (EG) | 1113.2 | 0.0161 | 0.0484 |

Though the proposed modification to the VoF method, expressed in Eqs. (9) and (10), was derived based on an experimental study of drop impact on thin films, its veracity and general applicability has been tested with other two-phase flow problems as discussed later. An example of the significance of property estimation equation is shown in Figure *2*. The plot shows the variation in interface density in the x-axis, between air ($\rho = 1.225$ kg/m$^3$) and a viscous liquid (ethylene glycol, $\rho = 1113.2$ kg/m$^3$). As the phase fraction ($\alpha$) in the x-axis goes from 0 to 1, the interface moves from air to that of the viscous liquid. As seen, from the two plots, the change in density with the exponential averaging equation (red line) is not monotonically linear as is the case with arithmetic averaging (green line). This ensures a sharper interface that is not smeared over multiple cells, thereby providing better estimates of the interface curvature and therefore, surface tension force and interface pressure drop.

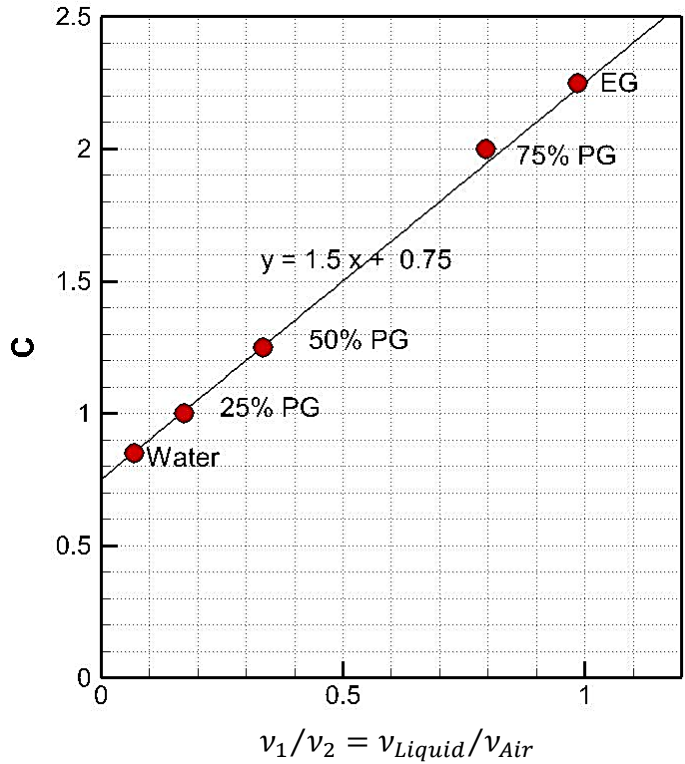

where,

**x % PG**: **x** % by volume of propylene glycol and water mixtures

EG: ethylene glycol

Figure 1: Equation for the determination of constant *C* in the property averaging expression I Eq. (9), as a function of relative viscosity ratios and as obtained from the mesurements with different air-liquid systems [note: EG – ethylene glycol, and x% PG – x% by volume propylene glcycol and water solution].

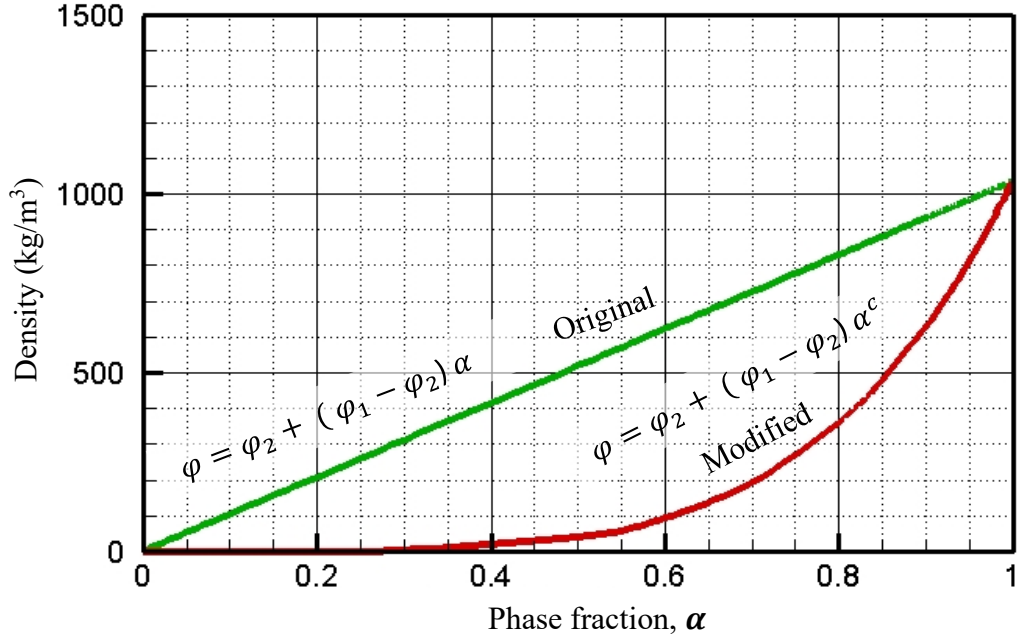

Figure 2: Comparison of the variation of fluid property (density) at the interface given by the arithmetic weighted averaging (green line) method of Eq. (2) and that by the new proposed method (red line) given by Eq. (10).

In addition to altering the property estimation, the curvature field is smoothened in accordance with the Ubbink et al. [20] technique where the phase fraction smoothness is corrected to influence the smoothness of the interface curvature, as the two are related by Eq. (*7*). This involves averaging the phase fraction in each computational cell volume by area averaging with respect to the values at each face as expressed in the following:

$$\alpha[cell] = \frac{\sum \alpha[face]\, Area[face]}{\sum Area[face]} \qquad (11)$$

Thus, the final set of equations that are solved in the proposed new modified VoF method are the continuity, momentum and phase transport equations along with the surface tension force determined by Eq. (*8*). The properties in this one fluid numerical approach (VoF) are estimated as per Eq. (*10*), and the interface curvature is further smoothened by area averaged smoothing of the phase fraction as given by Eq. (*11*).

**Solution Procedure**

The equations described in the previous section are discretized by using upwind scheme for the spatial terms and solved iteratively on the computational grid. From the computed variables the average values are stored in the cell center. Reconstruction from these averaged values is required to evaluate the cell interface values which need to be used for the subsequent iteration in the solution process. Limited linear piecewise reconstruction is used to diminish large variations during this reconstruction. The PIMPLE algorithm (combination of PISO and SIMPLE) is used as the iterative procedure for coupling mass and momentum conservation. For this coupling, within each time solution, this algorithm solves the pressure equation while invoking an explicit correction to velocity. Optionally, each iteration step can begin with a solution to the momentum equation. This is called the momentum predictor loop and is set to loop twice for each step. The linear solvers used in the PIMPLE method are *preconditioned Conjugate Gradient* for pressure and a *preconditioned Bi-Conjugate Gradient* for velocity components.

Because a transient solution is the goal of these numerical solutions, an implicit Euler time discretization is used. This first order implicit scheme is preferred to an explicit scheme as this guarantees boundedness and is unconditionally stable. For providing good numerical stability, adaptive time stepping is used where at the beginning of each time loop, the time step $\Delta t$ is calculated dynamically using the following criteria:

$$\Delta t = \min\left\{\frac{\text{Co}_{max}}{\text{Co}^o}\Delta t^o;\ \left(1 + \lambda_1 \frac{\text{Co}_{max}}{\text{Co}^o}\right)\Delta t^o;\ \lambda_2 \Delta t^o;\ \Delta t_{max}\right\} \qquad (12)$$

In the above criteria, $\Delta t_{max}$ and $\text{Co}_{max}$ are predefined values for maximum permissible time step and maximum Courant number. The damping factors, $\lambda_1$ and, $\lambda_2$, are defined as 0.1 and 1.2 respectively. The superscript *o* refers to values at the previous time-step. The maximum Courant number is set to be 0.1 for the current numerical study.

**Numerical domain and mesh discretization**

The numerical problems used in this study to test and establish the general applicability of the modified VoF for two-phase flows are all considered to be axisymmetric. This assumption is based on experimental evidence and helps reduce the computational domain and therefore makes the numerical solution faster. The entire physical domain of the problem can be considered as cylindrical with the axis representing the axis of symmetry. In OpenFOAM, the computational domain used for these problems is a wedge-shaped thin sliver (with a small angle, < 5º) of this cylinder. This wedge straddles the X-Y plane and runs along the central axis of symmetry as shown in Figure *3* (a). The typical orthogonal structured non-uniform mesh for this domain is shown in Figure *3* (b). The flow phenomena under examination are typically around the axis of symmetry in all the problems considered. Therefore, the mesh is denser in this region than farther away. Likewise, the mesh is compact near the target surface (for specific cases discussed where the fluid interacts with a target surface), to ensure accurate capture of the phenomena in these regions. Prior to obtaining the presented numerical results, the density of the grid is varied until a grid independent solution is obtained. For drop impact in a liquid pool, penetration depth was the parameter used to compare different meshes for interFoam method for 5cSt liquid. In the liquid jet breakup in stagnant ambience, breakup length of the continuous jet was the parameter used to compare different meshes for InterFoam method. The same mesh density was used for other cases studied. In modeling drop impact on thin liquid film, crown diameter was the parameter used to compare different mesh sizes for InterFoam to obtain grid-independent solutions.

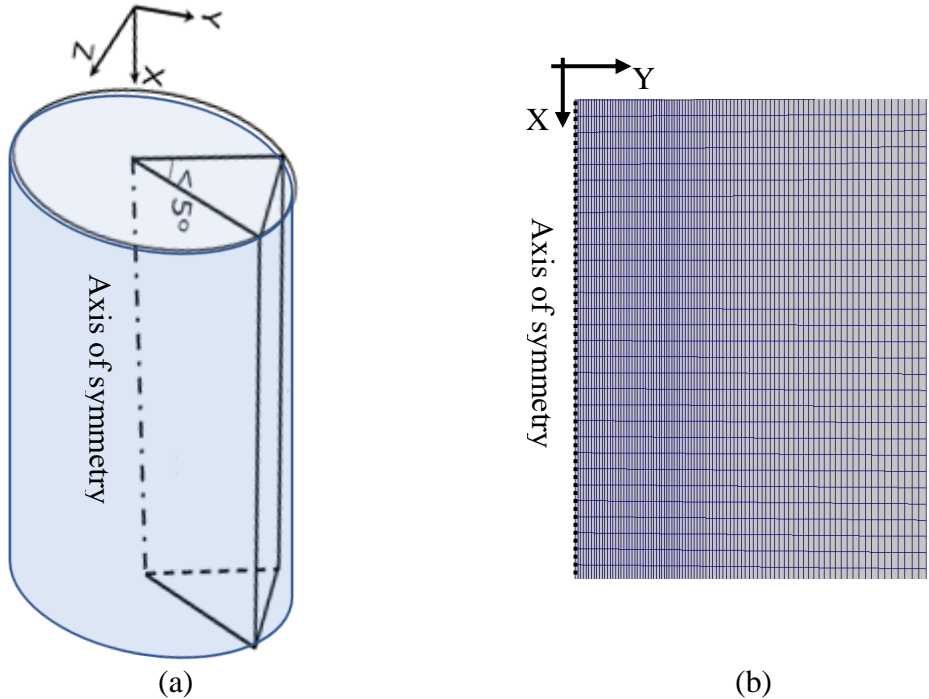

(a)　　　　　　　　　　　　　　　　　(b)

Figure 3: (a) the wedge-shaped numerical domain (b) typical mesh structure

**Performance tests on modified VoF method**

All implicit multi-fluid numerical methods rely on interface capturing techniques that are known to be susceptible to numerical instabilities and can result in unrealistic interface flows. When the flow phenomena under consideration are dominated by inertia, these numerically generated interface flows do not lead to unrealistic results. However, when capillary effects (surface tension forces) are dominant, any errors in calculating interfacial flows can give rise to inaccurate solutions. To compare and test the efficiency of the two numerical VoF solvers, namely, *interFoam* and the modified solver, four tests problems are solved and the results obtained are outlined in the next section. These include the canonical suspended drop problem and three other cases typically encountered in engineering applications (jet breakup, drop impact on a thin film, and drop impact on a liquid pool). In all these cases, surface tension forces are significant and that helps identify the difference in the accuracy of the two methods.

**Stationary drop in stagnant zero-gravity ambience**

The native VoF scheme (in interFoam solver), yields an interface curvature calculation with large gradients, thus contributing to a relatively larger spread of the interface. Consequently, the error in curvature results in an imbalance between the pressure and surface tension forces that lead to formation of spurious currents in the domain. The proposed modified VoF solver aims at creating a sharper interface and thereby improves the curvature estimation. To evaluate the influence of the modified treatment of interface, a stationary axi-symmetric drop in absence of gravity in a medium of air is studied. The initial and boundary conditions for this numerical analysis are shown in Figure *4*. A 4-mm diameter liquid drop with surface tension of 72.8 mN/m and density 1000 kg/m$^3$ is placed in stagnant air. The domain has a uniform grid distribution with a grid spacing of 0.001 mm. Two different drop viscosities (0.001 Pa.s and 0.01 Pa.s) are tested to see the influence of viscosity on the numerical generated current. These currents, called parasitic currents appear as vortices around the interface and, as shown in Figure *5*, are present when there are no external forces acting in the physical domain. This numerical error is not eliminated by grid refinement. In fact, as noted by Brackbill [22], the magnitude of these spurious currents could get amplified by a finer grid. Therefore, for unbiased comparison between the two methods, the same grid size and time stepping were used to check magnitudes of parasitic currents.

Figure *5* shows the spurious currents around a stationary drop in air in the absence of gravity. The observed currents for both *interFoam* and the modified VoF solver is shown for water and a more viscous (10 times that of water) liquid drop. It is observed increasing viscosity of the liquid, increases the intensity of the parasitic currents around interface of the stationary drop. The results shown in Fig. 5 are at 200 ms. The typical velocities in interFOAM are ~ 0.1 m/s while those with the modified solver at a tenth of that. The modified VoF solver, does not eliminate these currents, but the intensity of the same is noted to be greatly reduced. Thus, it is noted that a more accurate sharper approximation of the interface in the modified VoF solver can help reduce the spurious currents in the two-phase problem. To estimate the influence of these numerically induced currents on the domain, the kinetic energy around the drop interface is presented in Figure *6*. As is observed, the kinetic energy at the interface of this viscous drop in the numerical solution obtained by the modified VoF solver is an order of magnitude lower than that given by the *interFoam* solver.

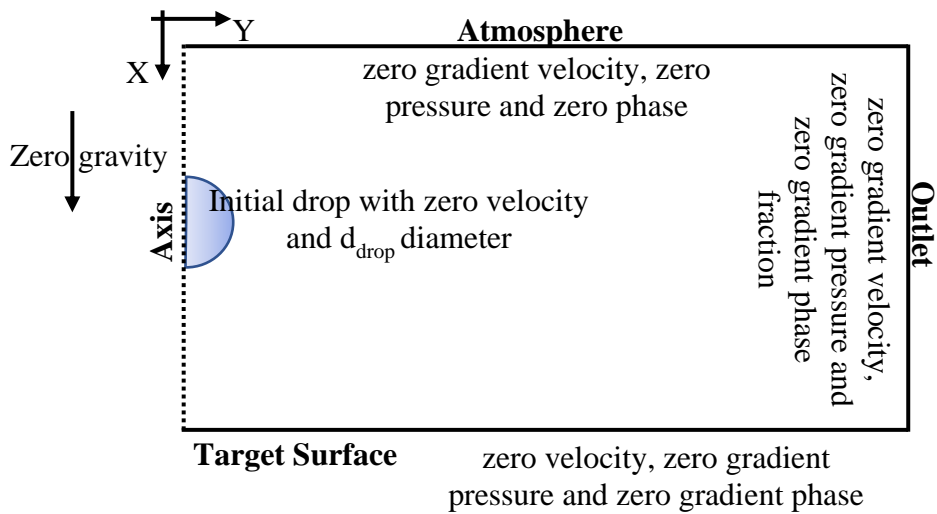

Figure 4: Initial and boundary conditions for numerical solution of stationary drop under no external forces.

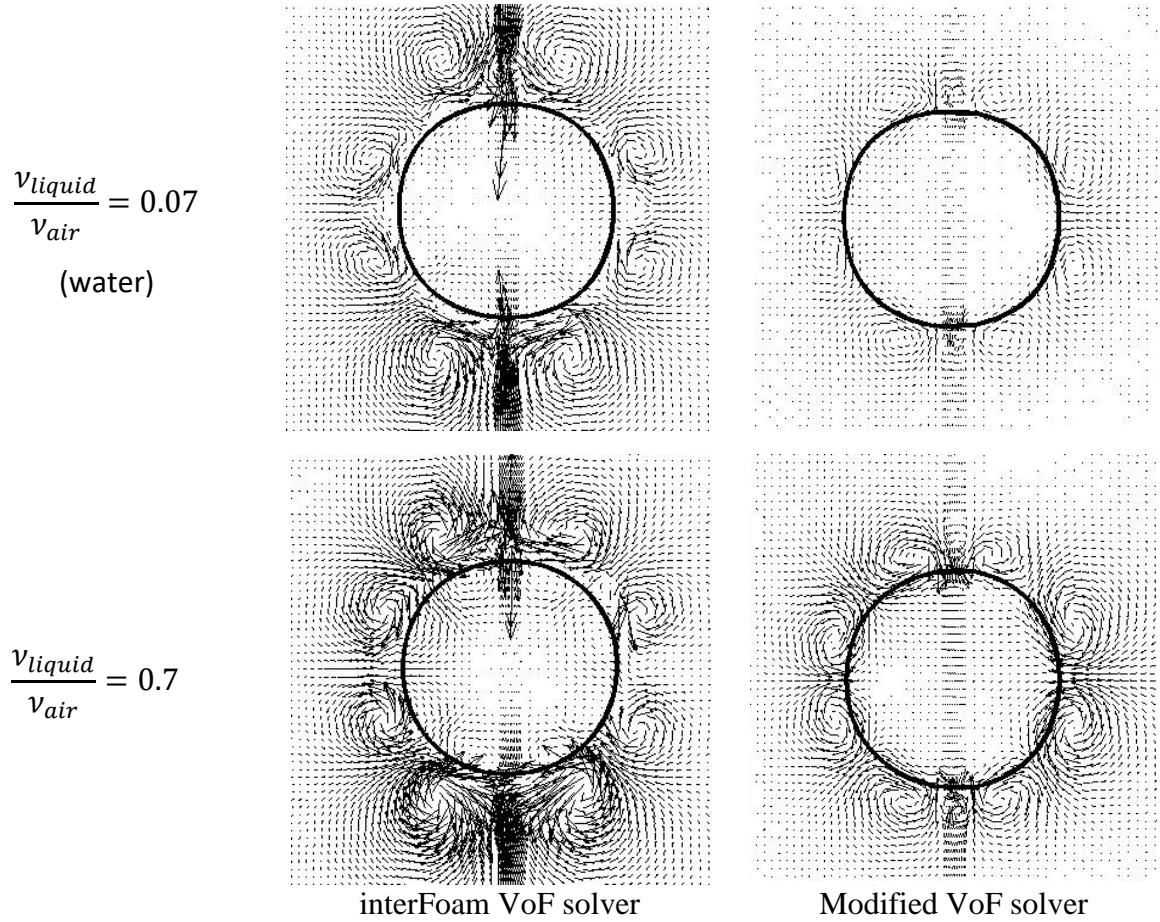

Figure 5: Comparison of the two VoF solvers: spurious currents around the interface of a stationary drop in ambient air in the absence of gravity at simulation time of 200 ms.

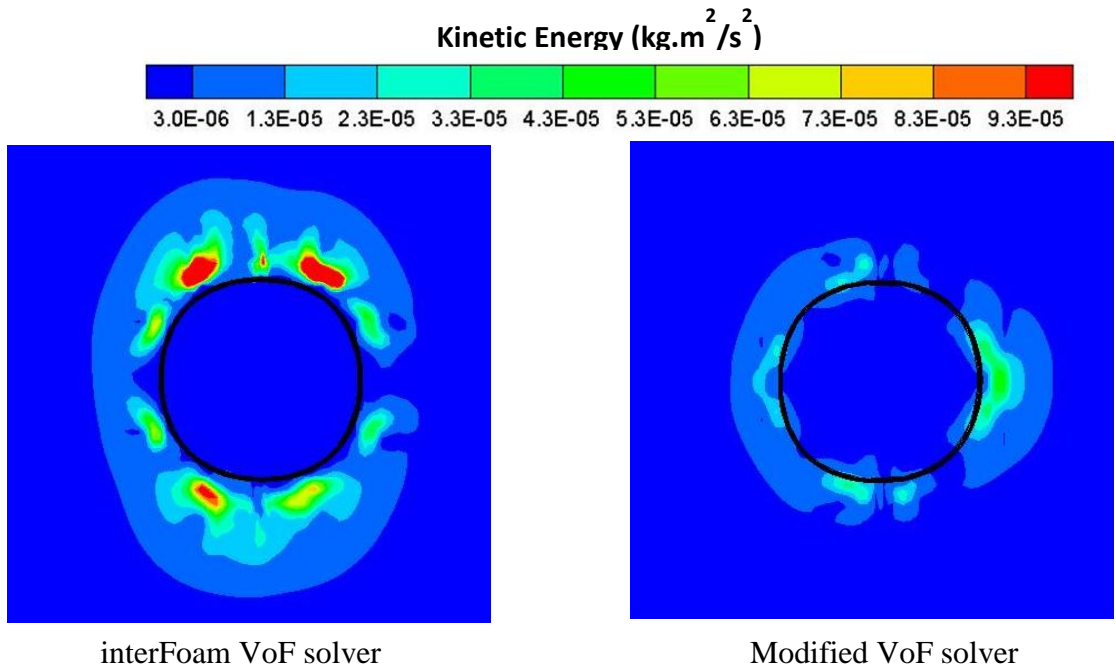

| interFoam VoF solver | Modified VoF solver |

Figure 6: Comparing kinetic energy around the viscous $\left(\frac{v_{liquid}}{v_{air}} = 0.7\right)$ stationary drop interface for interFoam and the modified VoF solver

**Drop impact in a liquid pool**

For the second test, experimental findings of Tran et al. [27] on the impact of a viscous drop on a deep liquid pool with attention to the entrapped air layer is considered. The initial temporal dynamics is studied for these impact conditions with the different viscous liquids used in the experimental work. At low impacting velocities, the air layer underneath the impacting drop prevents merger of the drop and pool surface and the drop could bounce off [28].Increasing the impact velocity was observed to aid entrapment of the air layer. The centerline depth as the drop merges with the pool (penetration depth) was studied as a function of time taken for rupture and was found to be linearly increasing with time. In addition, viscosity was noted to delay rupture time as higher the viscosity, more the resistance to allow entrapped air to rise and rupture the interface. These experimental findings are presented in Figure 7.

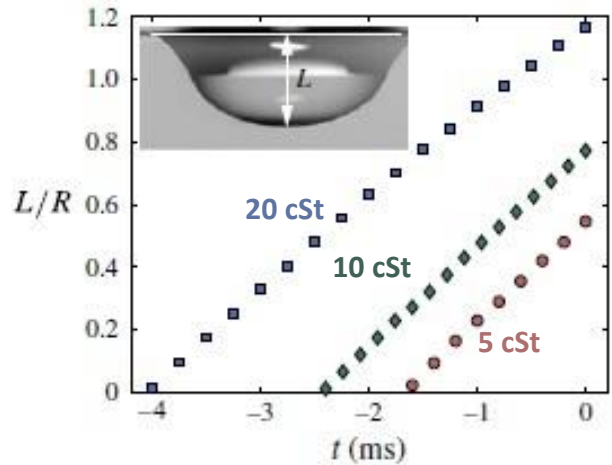

Figure 7: Experimental data showing dimensionless penetration depth (L/R) as a function of time of impact of drops with same impacting velocity (0.55 m/s) for 3 different viscosities [27]

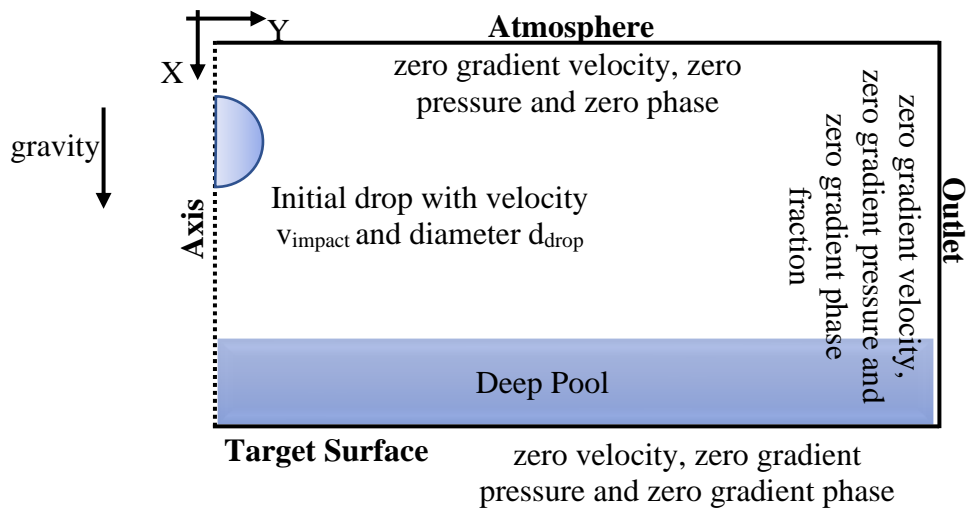

Figure 8: Initial and boundary conditions for numerical solution of drop impact on a deep pool.

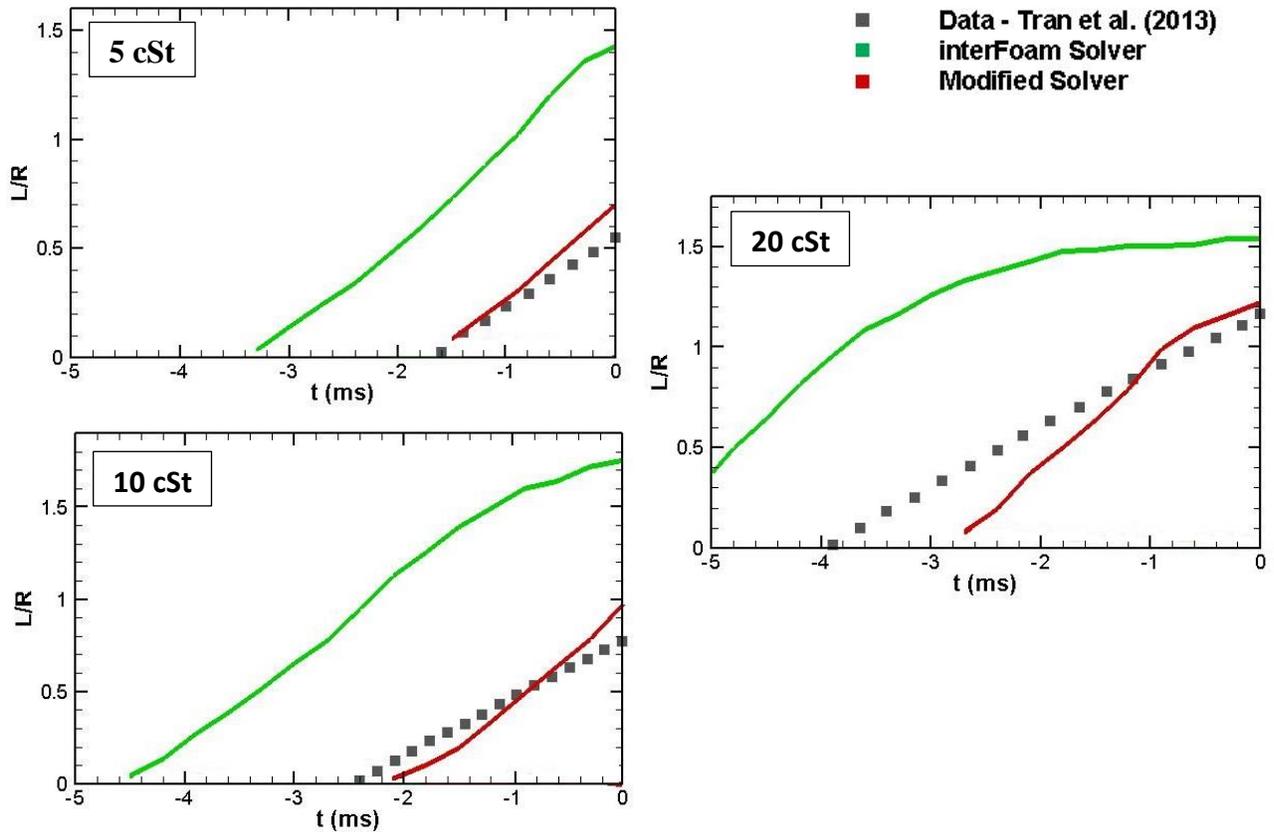

Figure 9: Performance of *interFoam* and modified solver against experimental findings [27] for entrapped air rupture time during viscous drop impact on a deep pool

To test the effectiveness of the modified VoF solver against the data, a drop impact on a pool in similar conditions is numerically simulated using both solvers (interFoam and modified VoF). The initial and boundary conditions for this problem are shown in the schematic in Figure *8*. A drop with diameter 1.9 mm and velocity of 0.55 m/s as provided by Tran et al. [27] impacts a pool of like liquid (10 mm deep) as shown in Figure *8*. The mesh for this domain under consideration contains approximately 390000 cells. While analyzing the effectiveness of the two solvers, it is observed that interFoam predicts a much slower evolution of air entrapment and rupture than the modified solver. The findings from both the solvers are compared with experimental data provided by Tran et al. [27] in Figure *9*. For all the three viscosities tested, the modified VoF solver is better at predicating the entrapped air dynamics. At lower viscosities (5 cSt), the modified solver is more precise at predicting the development of penetration depth, than at higher viscosities (20 cSt) compared to experimental measurements. For all the three viscosities, penetration depth at rupture ($\tau = 0$) are comparable with experimental data. Comparisons of dimensionless penetration depth for the three liquids for the two solvers is shown in Table 2.

*Table 2: Comparing dimensionless penetration depths for the three cases in consideration against interFoam and the modified VoF solvers.*

| Liquid's viscosity (cSt) | L/R expt. | L/R interFoam | L/R Modified | % difference (interFoam) | % difference (Modified) |
|---|---|---|---|---|---|
| 5 | 0.54 | 1.43 | 0.68 | 165 | 26 |
| 10 | 0.77 | 1.75 | 0.92 | 127 | 19 |
| 20 | 1.16 | 1.55 | 1.22 | 34 | 5 |

While the development time for the rupture is very comparable for 5cSt and 10 cSt drops, for the 20 cSt drop impact, total time for rupture is shorter than experimentally measured. As observed in the previous case study with numerically induced spurious currents in stationary drops, higher viscosity gives rise to stronger parasitic currents. Thus, with increasing viscosity, there is relatively more error in predicting the pressure and surface tension balance and thus the interface. It is speculated that this causes the numerical analysis to be more accurate when viscosities are lower. However, comparing with the experimental data, the deviation is significantly less than the original VoF solver.

**Liquid jet breakup in stagnant ambience**

Precise numerical characterization of capillary driven phenomena like liquid jet breakup requires correct estimation of surface tension and pressure forces on the interface. This becomes more significant when low velocity liquid jets are taken into consideration. A circular jet breaking up in stagnant ambience is governed by the interplay of inertial, gravitational, aerodynamic, capillary, and viscous forces. When the jet is fast moving, though surface tension (capillary force) is a prominent cause for creation and propagation of surface instabilities, inertial and aerodynamic forces are most dominant in the interplay [29–31]. This balance however changes when the jet is slow moving (smaller, lower weber numbers).

To investigate the accuracy of the two solvers in capturing the temporal growth of the surface instabilities, experimental data from our study of Newtonian low velocity viscous liquid jet breakup is used [32]. A propylene glycol jet of initial diameter 1.499 mm emerges from a circular nozzle at We = 5 and disintegrates to generate drops in stagnant ambience under the presence of gravity. Experimental observations show the phenomena to be symmetric about the axis and hence an axisymmetric wedge-shaped mesh with non-uniform structured mesh with about 160000 cells is used for this study. The initial and boundary conditions for this problem are shown in Figure *10*.

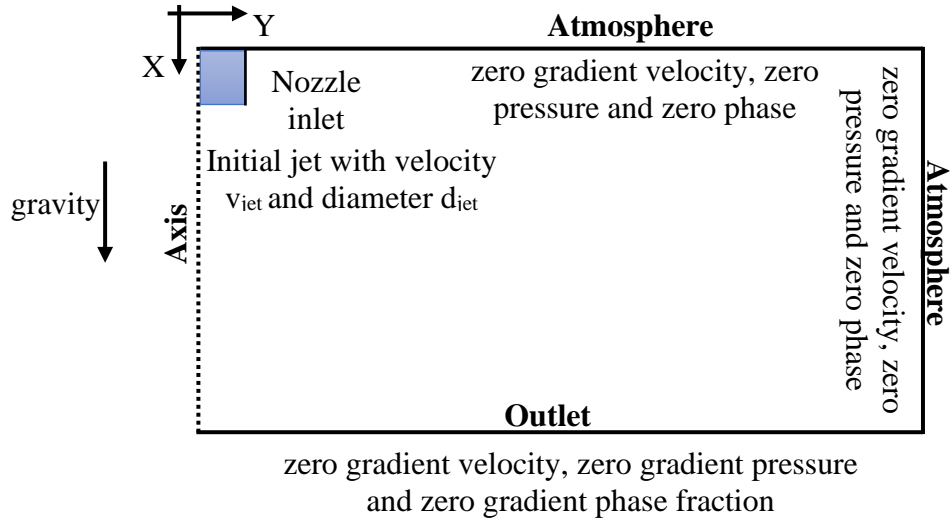

Figure 10: Initial and boundary conditions for numerical solution of jet breakup.

The nozzle wall has a no-slip boundary condition imposed on it. To emulate the experimental test conditions, the inlet has a uniform velocity inlet condition defined based on the prescribed inlet velocity from the syringe pump in the experimental study [32, 33]. At the outlets, the pressure is assumed to be atmospheric and zero-gauge pressure is the prescribed boundary condition. Breakup length calculations and other analyses of results are carried out after 0.2 s of real time. This provides time for numerical instabilities, if any, to deteriorate.

Figure *11* show the time progression of surface instability on the propylene glycol jet considered. High-speed experimental images are compared with predictions from the numerical solvers tested. It is observed that while the original solver (*interFoam*) is unable to capture the generation and propagation of surface instabilities unlike the modified VoF solver for high viscosity liquid jet. For water, this is however not the case and both solvers agree with the experimental observation. As propylene glycol is 40 times as viscous as water, and the VoF solver

predicts a much longer unperturbed jet surface compared to experimental observations. The modification on the property averaging helps improve the interface calculation and the modified solver can be seen to provide results that agree well with experimental data. With increasing viscosity of the external fluid, jet breakup exhibits different behavior (see for example, Borthakur et al. (2017) [ ]). The proposed method will be useful in addressing these situations as well.

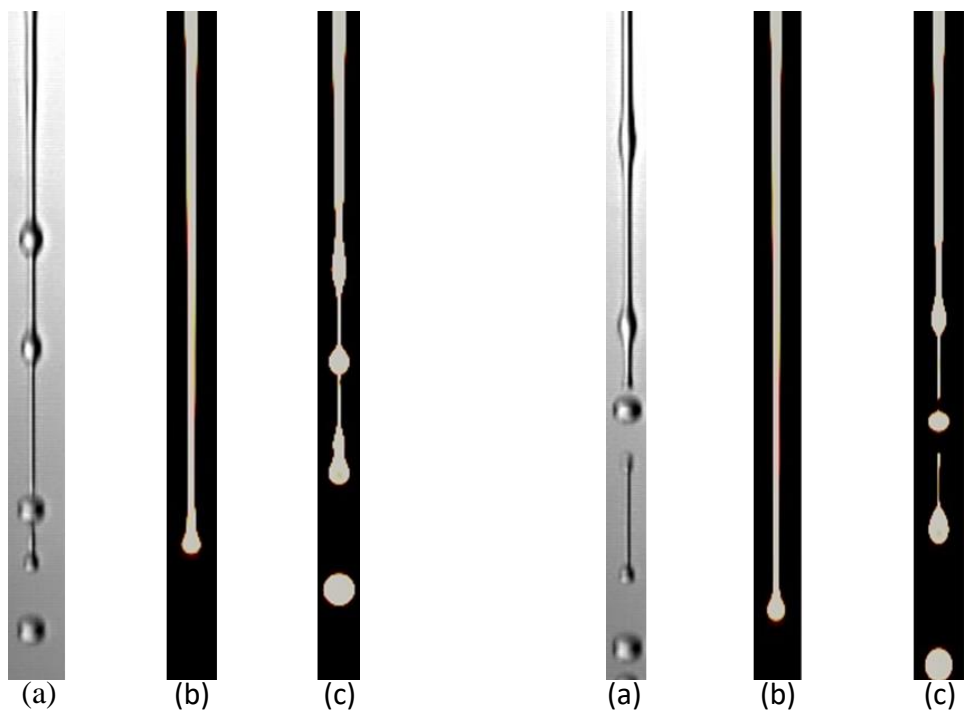

(a)    (b)    (c)      (a)    (b)    (c)

Figure 11: Comparison between (a) experimental [32, 33], and numerical predictions of (b) interFoam and (c) modified VoF for evolution of surface instabilities during circular jet breakup of a propylene glycol jet of 1.499 mm diameter and We = 5. The time difference between these two images is 2.5 ms.

**Drop impact on thin liquid film**

During drop impact on a thin liquid film of the same liquid, viscosity and surface tension affect the spread on the thin film and the crown growth. Understanding the correct interplay of these two forces can give insights into the growth and spread of the impacting drop as well as help enhance or curb splashing as per the requirement of the application such as coating and pesticide dispersions. A wedge-shaped axisymmetric domain (Figure *3* (a)) with structured mesh is used for numerical analysis of this problem. In the stagnant ambience, the drop of liquid with known impact velocity and diameter (from experimental observations) are initialized along with the stagnant thin film height and a schematic of the same is shown in Figure *12*. Figure *13* shows the time progressions of an ethylene glycol drop of diameter 4.56 mm impacting at a velocity of 1.89 m/s on a thin film. Successive images are at a gap of 2 ms and the temporal progression of the impact dynamics as predicted by the two solvers are compared with the experimental observations reported in [33]. While the initial spread of the drop on the film surface looks similar (at 2 ms and 4 ms), the crown tips are noticed to be thinner in the numerical solution using interFoam. This thinner crown tip leads to a faster moving crown rim that predicts the pinch-off secondary drop from the crown rim at 6, 8 and 10 ms after the impact. The modified VoF solver is able to predict the correct temporal and spatial dynamics of the drop liquid interaction as is seen in Figure *13*. Comparison of the spatial dynamics is done by measurement of the crown diameter as the drop merges with the thin film (at 10 ms in Figure *13*). The maximum difference between the solver predictions and that from the experiment for different Newtonian viscous liquids are compared in Table 3. The properties of the working liquid and the impacting drop are also provided in the table. The predictions using the modified solver are in agreement with both the final crown diameter as well as with the temporal development of the crown growth as shown in Figure *13*. We note that the regimes of drop impact dynamics discussed here, the experimental observations showcase a very symmetric development of the crown and breakup. This is the reason to consider axisymmetric 3-D simulation. This is similar to other studies in the literature studying similar drop impact dynamics (see for example, Ref. [3]).

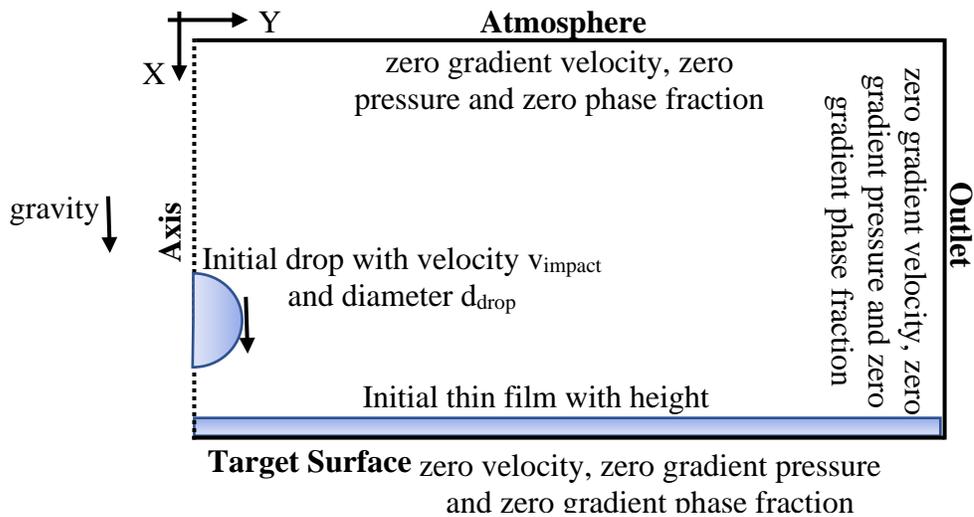

Figure 12: Initial and boundary conditions for numerical solution of drop impact dynamics.

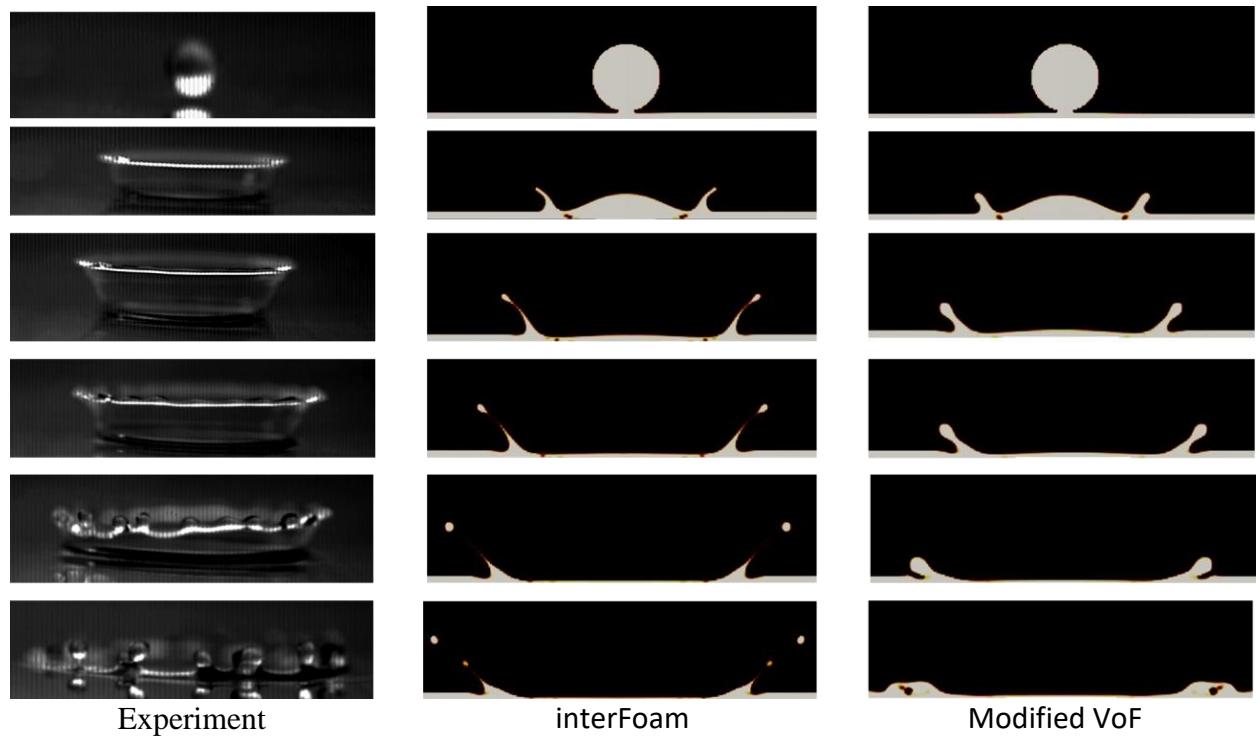

Figure 13: Comparison of numerical prediction from the two solvers with experimental time progression of drop impact dynamics (Successive image are 2 ms apart)

Table 3: Percentage difference between the experimental and numerical crown diameter

| Liquid | Density (kg/m$^3$) | Viscosity (Pa.s) | Surface Tension (mN/m) | $V_{impact}$ (m/s) | $D_{drop}$ (mm) | % difference (interFoam) | % difference (Modified) |
|---|---|---|---|---|---|---|---|
| Ethylene Glycol | 1113.2 | 0.0161 | 48.4 | 1.89 | 4.56 | 7.67 | 1.26 |
| 25% Propylene Glycol and Water | 1007.5 | 0.00255 | 54.1 | 1.87 | 3.79 | 13.53 | 1.92 |
| 50% Propylene | 1017.0 | 0.005 | 45.2 | 1.91 | 4.68 | 9.61 | 0.89 |
| Glycol and Water | | | | | | | |
| 75% Propylene Glycol and Water | 1026.5 | 0.012 | 41.1 | 2.38 | 4.93 | 8.12 | 1.50 |

**Conclusions**

The volume of fluid or VoF method has been used for numerical analysis of many inertia dominated as well as surface tension dominated two-phase problems. However, the current study finds that these predictions of two-phase flow phenomena tend to deviate significantly from experimental data for liquids with high viscosity. A new exponential property averaging technique is proposed to address this shortcoming. Computational predictions using the open source VoF solver *interFoam* (available as a part of the OpenFOAM computational tool), and those obtained using the proposed method are compared with experimental data for multiple two-phase applications. Four different problems, viz., suspended droplet in air, jet breakup, drop impact on thin films, and air entrapment during drop interaction with liquid pool, are considered to extensively validate the new method. For all cases, the modified VoF solver is observed to perform significantly better than original VoF method. For the suspended drop problem, as has been observed in previous numerical studies, spurious currents are noted to be a key source of numerical error. The proposed method reduces any spurious currents in simulations of drop suspended in air. For the cases of drop impacting on a pool and during drop generation from liquid jets, the time progression of the surface tension governed dynamics is improved from the slower estimate of *interFoam* solver. In the case of drop impacting on a thin liquid film, where it is critical to capture the intricate interplay between the surface tension and viscous force, the former effect of surface tension is exaggerated in the original method but correctly captured with the proposed new method.